\documentclass[fleqn,twoside]{article}
\usepackage{espcrc3}
\readRCS
$Id: espcrc2.tex,v 1.2 2004/02/24 11:22:11 spepping Exp $
\usepackage{graphicx}
\usepackage{epsf}

\newcommand{\AmS}{{\protect\the\textfont2
  A\kern-.1667em\lower.5ex\hbox{M}\kern-.125emS}}

\newbox\ourfigbox
\def\SizedFigureWithCaption#1#2#3{%
\setbox\ourfigbox
  \hbox{\epsfysize #1 \epsfbox{#2}}
\hbox to \wd\ourfigbox{\vbox{\noindent\copy\ourfigbox\break
\vskip -6mm      \hbox to \wd\ourfigbox{\hss#3\hss}}}
}

\def\e{\epsilon}

\def\tree{{(0)}}
\def\oneloop{{(1)}}
\def\twoloop{{(2)}}

\def\Lloop{{(L)}}
\def\lloop{{(l)}}
\def\Llloop{{(L-l)}}
\def\Ord{{\cal O}}
\def\Prop{D}

\def\Split{\mathop{\rm Split}\nolimits}

\def\n{n}    

\def\rsn{r_S}

\def\Neqfour{{{\cal N}=4}}

\def\fig#1{fig.~{\ref{#1}}}

\def\eqn#1{eq.~(\ref{#1})}

\title{Two-loop Splitting Amplitudes}

\author{Z. Bern\address[MCSD]{Department of Physics and Astronomy, UCLA \\
             Los Angeles, CA 90095-1547, USA}\thanks{Presenter at 
        7th DESY Workshop on Elementary Particle Theory, Loops and Legs in 
         Quantum Field Theory, April 25-30, Zinnowitz (Usedom Island), 
         Germany. 
         Research supported by the US Department of
        Energy under contract DE-FG03-91ER40662.},
        L.J. Dixon\address[MCSD]{Stanford Linear Accelerator Center, 
          Stanford University\\
                  Stanford, CA 94309, USA}\thanks{Research supported 
             by the US Department of Energy under contract DE-AC03-76SF00515.} 
        and 
        D.A. Kosower\address{ Service de Physique Th\'eorique, CEA--Saclay\\
        F-91191 Gif-sur-Yvette cedex, France}
           }
       
\runtitle{Two-Loop Splitting Amplitudes}

\begin{document}

\begin{abstract}

Splitting amplitudes govern the behavior of scattering amplitudes at
the momenta of external legs become collinear.  In this talk we
outline the calculation of two-loop splitting amplitudes via the unitarity
sewing method.  This method retains the simple factorization
properties of light-cone gauge, but avoids the need for
prescriptions such as the principal value or Mandelstam-Leibbrandt ones. The
encountered loop momentum integrals are then evaluated using
integration-by-parts and Lorentz invariance identities.  We outline a
variety of applications for these splitting amplitudes.
\vskip -.1 cm 
\end{abstract}

\maketitle

\section{INTRODUCTION}

Recent years have seen rapid progress in computing higher-order
corrections to the standard model. In particular, there have been a
large number of new computations of two-loop scattering amplitudes
with more than a single kinematic variable (see
refs.~\cite{TwoloopCalculationReview} and references therein.).  In
the past decades, a number of new approaches have been developed to
cope with this complexity, including helicity methods, color
decompositions, recursion relations, ideas based on string theory, and
the unitarity-based method, summarized in a variety of review
articles~\cite{MPReview}.  Much of the progress at two loops has been
facilitated by new techniques for performing loop integrals (see
refs.~\cite{TwoloopIntegralsReview}).

Splitting amplitudes describe the universal singular behavior of
amplitudes in the regions of phase space where momenta become
collinear.  In this talk we summarize the calculation of $g
\rightarrow gg$ splitting amplitudes at two loops via the unitarity
sewing method~\cite{Split2}. The two-loop splitting amplitudes,
including the ones involving quarks, have also been extracted recently
from explicit computations of four-point helicity amplitudes with a
massive leg~\cite{SplitGlover}.  With the unitarity sewing method the
simple factorization properties of light-cone gauge are retained, but
the need for principal value or other such
prescriptions~\cite{Leibbrandt} is avoided.

We also discuss a variety of applications of two-loop splitting
amplitudes.  These include verifying amplitude
calculations~\cite{MPReview} or constructing
ans\"atze~\cite{ParkeTaylor,Neq4Oneloop} for amplitudes with more than
two kinematic variables. Splitting amplitudes also enter into an
alternative method~\cite{KosowerUwerKernel} for computing corrections
to the DGLAP kernel governing the $Q^2$ evolution of parton
distributions and fragmentation functions.  We have also used them in
a proof~\cite{Split2} of Catani's formula~\cite{CataniTwoloop} for
universal two-loop infrared divergences, complementing the proof based
on resummation~\cite{StermanIR}.  They have also been used to provide
direct evidence that in maximally supersymmetric $\Neqfour$
super-Yang-Mills theory, higher-loop planar amplitudes can be
expressed in terms of lower-loop ones, suggesting that substantial
portions of the theory may be solvable.

\section{SPLITTING AMPLITUDES}

\begin{figure}[tb]
\centerline{\epsfxsize 2.9 truein \epsfbox{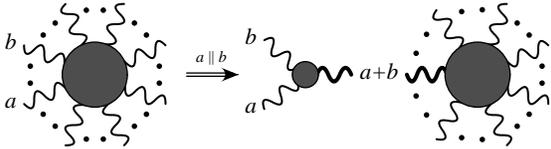}}
\vskip -.7 cm 
\caption{The collinear factorization of a tree-level amplitude.
The thick line represents a slightly off-shell gluon.}
\label{CollinearTreeFigure}
\vskip -.3 cm 
\end{figure}

For simplicity, here we consider only the leading color contributions.
The full color structure of the two-loop splitting amplitudes may be
found in refs.~\cite{Split2,SplitGlover}.  At $L$ loops the behavior
of the leading-color contributions are given
by~\cite{MPReview,ParkeTaylor,Neq4Oneloop,OneloopSplit,KosowerUwerSplit}
\begin{eqnarray}
&&  \hskip -.7 cm 
A_{n}^{\Lloop}(\ldots,a^{\lambda_a},b^{\lambda_b},\ldots) 
\mathop{\longrightarrow}^{a \parallel b}
\nonumber \\
&&  \hskip -.7 cm 
 \sum_{l=0}^L \sum_{\lambda=\pm}
  \Split^{\lloop}_{-\lambda}(z; a^{\lambda_a},b^{\lambda_b})\,
      A_{n-1}^{\Llloop}(\ldots,P^\lambda,\ldots) \,, \nonumber
\end{eqnarray}
in the limit where the momenta $k_a \rightarrow z k_P$ and $k_b
\rightarrow (1-z) k_P$ with $k_P = k_a + k_b$. Here
$\Split^{\lloop}_{-\lambda}(z;a^{\lambda_a},b^{\lambda_b})$ is an
$l$-loop splitting amplitude.  Legs $a$ and $b$ carry helicities
$\lambda_a$ and $\lambda_b$, while the merged leg $P$ carries helicity
$\lambda$, runs over the two helicities of the intermediate state. In
the other sum, $l$ runs over the loop order.  As a simple example, the
collinear factorization of an $n$-point tree amplitude into a
splitting amplitude and an $(n-1)$-point amplitude is depicted
schematically in \fig{CollinearTreeFigure}.

\begin{figure}[tb]
\centerline{\epsfxsize .9 truein \epsfbox{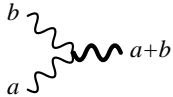}}
\vskip -.8 cm 
\caption{The three-point vertex diagram for obtaining the tree-level
splitting amplitude.}
\vskip -.3 cm 
\label{TreeThreeVertexFigure}
\end{figure}

At tree level, the splitting amplitudes are rather straightforward to
obtain by evaluating diagrams. The splitting amplitude depicted in
\fig{CollinearTreeFigure} may be computed directly from the Feynman
three-vertex depicted in \fig{TreeThreeVertexFigure} by multiplying by
the off-shell propagator and saturating the external legs with
helicity states~\cite{MPReview}.

Beyond tree level the analogous computation is more subtle.  In
covariant gauges there are additional `non-factorizing' contributions
associated with soft-gluon emission, complicating the
analysis~\cite{BernChalmers}.  One might be tempted to avoid this
difficulty by using instead physical gauges such as light-cone gauge
since they have simple factorization properties.  Unfortunately,
light-cone gauge has another set of complications arising from the
propagator,
\begin{eqnarray}
&& \Prop_{\mu \nu} = -{i\over p^2 + i \e}\, 
\biggl[\eta_{\mu \nu} - 
{n_\mu p_\nu + p_\mu \n_\nu \over p \cdot n } \biggr] \,,
\nonumber
\end{eqnarray}
where $p$ is the particle momentum, $\eta_{\mu\nu}$ is the Minkowski
metric, and $n$ is a null vector ($n^2 = 0$) defining the light-cone
direction.  The problem occurs in the region where $p \cdot n$
vanishes.  One common method for dealing with this singularity is the
principal-value prescription which replaces
\begin{eqnarray}
&& {1\over p \cdot n} \rightarrow  \lim_{\delta \rightarrow 0} 
 {1\over 2} \biggl( {1\over  p \cdot n + i \delta }
                  + {1\over  p \cdot n - i \delta } \biggr) \,,
\nonumber
\end{eqnarray}
where $\delta$ is a regulator parameter.  Another choice, better
founded in field theory, is the Mandelstam-Leibbrandt
prescription~\cite{Leibbrandt}.  The introduction of any of these
prescriptions in a splitting amplitude calculation is problematic for
a number of reasons.  The most serious problem is that results retain
dependence on the regulator parameter $\delta$.  (In general these
cancel only after combining virtual and real emission contributions.)
A calculation of the splitting amplitudes that retains this dependence
cannot match the splitting amplitudes extracted from the collinear
limits of scattering amplitudes which are independent of such a
parameter.  Thus one would not obtain the desired results.

\begin{figure}[tb]
\centerline{\epsfxsize 1.8 truein \epsfbox{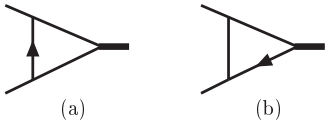}
}
\vskip -.9 cm 
\caption{One-loop triangle diagrams containing
light-cone denominators indicated by the arrow. Only integral
(b) appears in the unitarity sewing method.}
\label{OneLoopExampleFigure}
\vskip -.5 cm 
\end{figure}

\begin{figure}[tb]
\centerline{\epsfxsize 1.5 truein \epsfbox{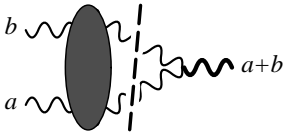}}
\vskip -.8 cm 
\caption{The two-particle cut of a one-loop splitting amplitude. The
cut is represented by the dashed line. On the left-hand side all legs,
including the cut ones, are on-shell.}
\label{OneLoopSplittingCutFigure}
\vskip -.4 cm 
\end{figure}

Is the need for these prescriptions is an artifact of the gauge
choice? Could they be avoided in a more physical construction of the
splitting amplitudes? Yes!  As a concrete example, consider the two
integrals in \fig{OneLoopExampleFigure}. In a light-cone gauge
calculation of the one-loop $g\rightarrow gg$ splitting amplitudes
both integrals appear.  What distinguishes these two integrals?  The
key is the unitarity cut shown in \fig{OneLoopSplittingCutFigure}.  On
the left-hand-side of the cut is an on-shell gauge invariant amplitude
which does not contain light-cone denominators. This leads one to
suspect that the integral in \fig{OneLoopSplittingCutFigure}(a) is
spurious and should not appear.  On the other hand, the light-cone
denominator in integral (b) does appear in the physical state
projectors of the unitarity cut suggesting that it cannot be removed.
Integral (b), however, is perfectly well defined using dimensional
regularization and is therefore harmless.

\section{CONSTRUCTION AND RESULTS}

To avoid the appearance of ill-defined integrals we use the unitarity
sewing method~\cite{Neq4Oneloop,Unitarity,KosowerUwerSplit,Split2}.
With this method the loop integrands are obtained algorithmically by
constructing unrestricted loop momentum integrals with the correct
cuts in all channels.  The method is well-tested, having been applied
to a variety of problems.  For example it has been used to obtain
analytic formulas for $Z \rightarrow 4$ partons at one
loop~\cite{Z4Partons} and one-loop maximally helicity violating (MHV)
amplitudes with an arbitrary number of external legs in supersymmetric
gauge theories~\cite{Neq4Oneloop}.  More relevant for the discussion
here, it has also been used to obtain one-loop splitting
amplitudes~\cite{KosowerUwerSplit}.

\begin{figure*}[htb]
\centerline{\hbox{%
\SizedFigureWithCaption{0.6 truein}{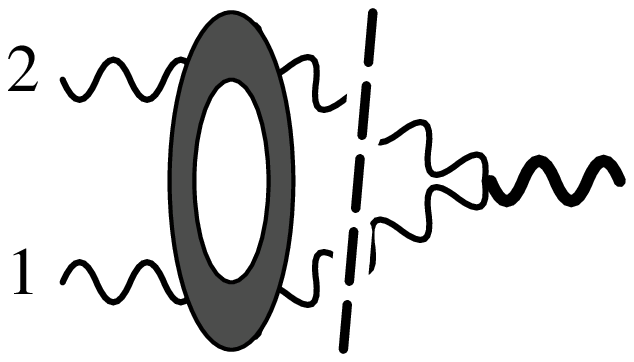}{(a)}
\hfil\hskip 2 truemm
\SizedFigureWithCaption{0.6 truein}{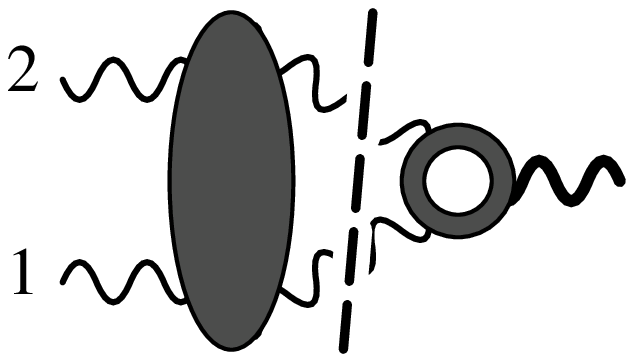}{(b)}
\hfil\hskip 2 truemm
\SizedFigureWithCaption{0.6 truein}{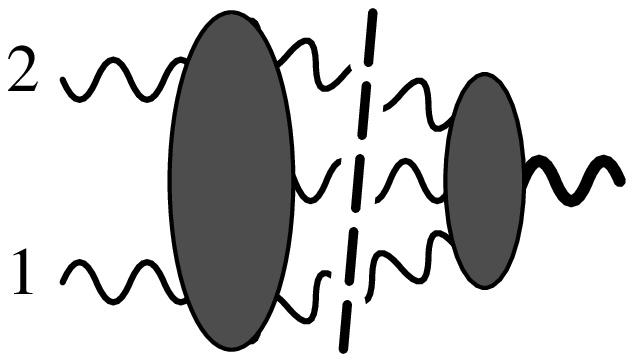}{(c)}%
}}
\vskip -.7 cm 
\caption{The three cuts contributing to the two-loop splitting
amplitude.}
\label{TwoLoopSplittingCutFigure}
\end{figure*}

The two-loop splitting amplitudes for $g \rightarrow gg$ were
constructed by combining the three contributing unitarity cuts shown
in \fig{TwoLoopSplittingCutFigure} into a single
integrand~\cite{Split2}.  One simplifying feature of the $g
\rightarrow gg$ splitting amplitudes is that the non-planar
configurations do not contribute, because of color considerations.
The resulting integrand was then evaluated in terms of 13 master
integrals using the implementation of the Laporta
algorithm~\cite{Laporta} due to Anastasiou and Lazopoulos~\cite{AIR}.
The master integrals were computed by constructing and solving a set
of differential equations, along the lines of
refs.~\cite{PBReduction}.

The results for the two-loop QCD splitting amplitudes are too lengthy
to include here, but are given in detail in ref.~\cite{Split2} where
they are expressed in terms in terms of a series expansion in $\e$,
with each term containing ordinary logarithms and polylogarithms.

The two-loop results for $\Neqfour$ super-Yang-Mills theory have an
especially intriguing feature.  To expose this, it is convenient to
first write the loop splitting amplitudes in terms of a ratio with the
corresponding tree functions,
\begin{eqnarray}
&& \hskip -.5 cm 
\Split_{-\lambda}^{(L)}(a^{\lambda_a},b^{\lambda_b}) 
\nonumber
\\
&& \hskip .5 cm
  = \rsn^{(L) \, \lambda_a,\lambda_b}(z,s_{ab}) \times 
      \Split_{-\lambda}^\tree(a^{\lambda_a},b^{\lambda_b}) \,,
\nonumber 
\end{eqnarray}
where $L$ indicates the loop order.  Rather surprisingly the two-loop
ratios are expressible in terms of the one-loop
ratios~\cite{TwoloopN4},
\begin{eqnarray}
&& \hskip -.5 cm 
r^{(2),\, \Neqfour}_S(\e;z, s) = 
   {1 \over 2} \Bigl( r^{(1),\, \Neqfour}_S(\e;z, s) \Bigr)^2 
\label{OneloopTwoloopSplit}
\\
&& \hskip 1 cm  \null  + f(\e)\, r^{(1),\, \Neqfour}_S(2\e;z, s)
     + \Ord(\e) 
\,.
\nonumber 
\end{eqnarray}
The explicit values of $r^{(1),\, \Neqfour}_S$, $r^{(2),\,
\Neqfour}_S$ and $f(\e)$ may be found in
refs.~\cite{TwoloopN4,Split2}.  This relation between the two-loop and
one-loop splitting amplitudes may provide a hint to unraveling the
theory.

\section{APPLICATIONS}

One use of splitting amplitudes is to provide checks on calculations
of amplitudes with more than two kinematic invariants from the
collinear constraints~\cite{MPReview}.  At two loops the $g
\rightarrow gg$ amplitudes are consistent with the collinear limits of
the two-loop $H\to ggg$ helicity
amplitudes~\cite{Koukoutsakis,Split2,SplitGlover}. In some cases the
collinear constraints are sufficient for constructing ans\"atze for
$n$-point scattering amplitudes~\cite{ParkeTaylor,Neq4Oneloop}.  It is
worth noting that after subtracting the divergences from an amplitude,
the finite remainders also satisfy universal collinear
constraints~\cite{Split2}, providing a direct check on these terms.

Another application of the two-loop splitting amplitudes is for a
proof of Catani's formula~\cite{CataniTwoloop} for two-loop infrared
divergences~\cite{Split2}. The proof is based on taking collinear
limit of the $n$-point formula, and comparing it with the
$(n-1)$-point formula.  The proof does require the reasonable
assumption that terms do not vanish in all collinear limits.  This
proof is complementary to the one based on
resummation~\cite{StermanIR}, since it fixes the explicit form of the
functions appearing in Catani's formula, including the previously
unknown $n$-point color non-trivial terms at order $1/\e$.

The two-loop splitting amplitudes are also one of the ingredients to
an alternative approach~\cite{KosowerUwerKernel} to computing the
DGLAP kernels at NNLO.  These kernels have recently been computed by
Moch, Vermaseren and Vogt~\cite{MVVNNLO} using the Mellin space
approach of computing anomalous dimensions of leading-twist operators.
The computation of these kernels are of great importance for precision
extraction of parton distribution functions from experimental data.
As discussed in ref.~\cite{Split2}, the splitting amplitudes can
straightforwardly be continued to the space-like region relevant for
the evolution of parton distribution functions.

A more theoretical application of splitting amplitudes is for
investigating the form of the multi-loop scattering amplitudes in
maximally supersymmetric Yang-Mills theory.  It has been 30 years
since 't~Hooft suggested that QCD could be solved in the planar limit.
Unfortunately, this has not been achieved as yet, so a logical
approach is to start with a simpler theory.  The structure of
maximally supersymmetric $\Neqfour$ Yang-Mills theory is much simpler
than QCD since it possesses superconformal symmetry.  Moreover, the
Maldacena conjecture~\cite{Maldacena} suggests that the strongly
coupled limit of the theory should be much simpler than one might
otherwise expect, since it is dual to weakly coupled gravity in
anti-de Sitter space.

The two-loop splitting amplitudes and their associated collinear
constraints lead to an ansatz for the planar contributions to the
$n$-point two-loop amplitudes in terms of the $n$-point one-loop
amplitudes of the form~\cite{TwoloopN4}
\begin{eqnarray}
&&M_n^{\twoloop}(\e) =  {1 \over 2} \Bigl(M_n^{\oneloop}(\e) \Bigr)^2
             + f(\e) \, M_n^{\oneloop}(2\e) \nonumber\\
&& \hskip 3.2 cm \null  - {5 \over 4} \,\zeta_4  +  \Ord(\e) \,,
\label{TwoloopOneloop}
\end{eqnarray}
where $f(\e)$ is the same function as appearing in
\eqn{OneloopTwoloopSplit}.  The $M_n^{\Lloop}(\e)$ are $n$-point
$L$-loop planar amplitudes, divided by the tree amplitudes.  For $n=4$
explicit calculation~\cite{BRY,SmirnovDoubleBox,TwoloopN4} confirms
this ansatz.  More generally, for $n\ge 5$ the ansatz
(\ref{TwoloopOneloop}) satisfies the correct collinear properties,
using the splitting amplitudes (\ref{OneloopTwoloopSplit}).  For the
maximally helicity-violating amplitudes it is likely that the ansatz
(\ref{TwoloopOneloop}) is correct, but for the non-maximally helicity
violating amplitudes it would be important to also check
multi-particle factorization.  The crucial question is whether this
iterative structure holds to higher loops and whether a resummation is
possible.  An important step to investigate the structure would be the
computation of the three-loop four-point amplitude.  This computation
seems feasible~\cite{SmirnovLoopsLegs}, especially since one of two
required integrals~\cite{BRY} has already been evaluated by
Smirnov~\cite{SmirnovThreeLoop}.


\end{document}